\documentclass{article}

\usepackage[dvips]{graphicx}

\def\abstract#1{\vskip 7mm 
        \begin{center}{\large Abstract}\par \smallskip
                \begin{minipage}[c]{12cm}
                        \small #1
                \end{minipage}
        \end{center}
}
\def\title#1{\begin{center}{\Large\bf #1}\end{center}}
\def\author#1{\vskip 5mm \begin{center}{#1}\end{center}}
\def\address#1{\begin{center}{\it #1}\end{center}}
\def\vx{{\bf x}}

\def\vk{{\bf k}}

\def\vp{{\bf p}}

\def\v0{{\bf 0}}

\makeatletter
\@ifundefined{lesssim}{}{}
\@ifundefined{gtrsim}{}{}
\def\vereq#1#2{\lower3pt\vbox{\baselineskip1.5pt \lineskip1.5pt
\ialign{$\m@th#1\hfill##\hfil$\crcr#2\crcr\sim\crcr}}}
\makeatother

\begin{document}

\title{% 
  Cancellation of long-range forces in Einstein-Maxwell-dilaton system
%  \smallskip \\
%  {\large --- Please use this file to complete your manuscript  ---}
}
\author{%
  Nahomi Kan$^*$\footnote{E-mail:kan@yamaguchi-jc.ac.jp},
  Koichiro Kobayashi$^\dagger$\footnote{E-mail:m004wa@yamaguchi-u.ac.jp},
  Teruki Hanada$^\dagger$\footnote{E-mail:k004wa@yamaguchi-u.ac.jp} 
  and
  Kiyoshi Shiraishi$^\dagger$\footnote{E-mail:shiraish@yamaguchi-u.ac.jp}
}
\address{%
  $^*$Yamaguchi Junior College,
   Hofu-shi, Yamaguchi 747--1232, Japan\\
  $^\dagger$Yamaguchi University,
  Yamaguchi-shi, Yamaguchi 753--8512, Japan
}

\abstract{
We examine cancellation of long-range forces in Einstein-Maxwell-Dilatonic system.
Several conditions of the equilibrium of two charged masses in general relativity
is found by many authors.
These conditions are altered by taking account of dilatonic field.
Under the new condition,
we show cancellation of $1/r^2$ potential
using Feynman diagrams.  
}

%%%%%%%%%%%%%%%%%%%%%%%%%%%%%%%%%%%%%%%%%%%%%%%%%%%%%%%%%%%%%%%%%%%%%%%%%%%
%Introduction
%%%%%%%%%%%%%%%%%%%%%%%%%%%%%%%%%%%%%%%%%%%%%%%%%%%%%%%%%%%%%%%%%%%%%%%%%%%
\section{Introduction}
The interaction between two charged massive particles is described by 
the Newton and  the Coulomb potential:
\begin{equation}
V(r)=-G\frac{Mm}{r}+\frac{1}{4\pi}\frac{Qq}{r},
\end{equation}
where $G$ is the Newton constant.
If $Qq=4\pi GMm$, $V(r)=0$
and long-range forces are canceled each other at the classical level.
In general relativity,
the static exact solution for the equilibrium of two charged masses 
was found by
Majumdar and Papapetrou~\cite{MP}.
The solution requires a condition:
$\sqrt{ 4 \pi G}M_i = Q_i$.
The static exact solution with dilaton was found by Shiraishi~\cite{KSJMP}.
The solution  stands for the case with the balance condition:
$(M_i:Q_i:\Sigma_i)=(1:\sqrt{1+a^2}:a)$, 
where $\Sigma$ is a dilatonic charge.
In this work,
we examine cancellation of long-range forces in the Einstein-Maxwell-Dilatonic system
using Feynman diagrams under the balance condition,
$Q=\sqrt{1+a^2}M$.
%%%%%%%%%%%%%%%%%%%%%%%%%%%%%%%%%%%%%%%%%%%%%%%%%%%%%%%%%%%%%%%%%%%%%%%%%%%
%Feynman rules for dilaton
%%%%%%%%%%%%%%%%%%%%%%%%%%%%%%%%%%%%%%%%%%%%%%%%%%%%%%%%%%%%%%%%%%%%%%%%%%%
\section{Feynman rules for dilaton}
We start with the Lagrangian including a dilaton field $\phi$:
\begin{equation}
\mathcal{L} =
\frac{\sqrt{-g}}{4}
(R-e^{-2a\phi}F^2+2g^{\mu\nu}\nabla_\mu\phi\nabla_\nu\phi)~,
\end{equation}
where $4\pi G=1$,
$R$ is Ricci scalar,
$F_{\mu\nu}=\partial_\mu A_\nu-\partial_\nu A_\mu$,
$\nabla_\mu\phi=\partial_\mu\phi$, and %, for a dilaton field $\phi$, and
$a$ is the coupling constant between the dilaton field and other fields.
%------------------------------------------------------------------------------
%In this study, examine, extract
In order to
extract the interactions coupled with the dilaton field, 
%evaluate the potential of long-range forces coupled with the dilaton field from Feynman diagrams,
%from ...
we use a complex scalar boson $\varphi$ as a probe.
The Lagrangian of complex Klein-Gordon field is~\cite{DS}
\begin{equation}
\mathcal{L}_{KG} = {\sqrt{-g}}\,[e^{-a\phi}g^{\mu\nu}(D_\mu\varphi)^*
D_\nu\varphi-m^2e^{a\phi}\varphi^*\varphi]~,  
\end{equation}
where
$D_\mu\varphi=\partial_\mu\varphi+iqA_\mu\varphi$.
%------------------------------------------------------------------------------
We decompose the metric $g_{\mu\nu}$ %as
into the flat background field $\eta_{\mu\nu}$
and the graviton field $h_{\mu\nu}$:
\begin{equation}
g_{\mu\nu}=\eta_{\mu\nu}+\kappa h_{\mu\nu},
\end{equation}
where $\kappa=\sqrt{32\pi G}$ and $\eta_{\mu\nu}=diag(1,-1,-1,-1)$.
%------------------------------------------------------------------------------
In this decomposition, we read 
\begin{equation}
g^{\mu\nu}=\eta^{\mu\nu}-\kappa h^{\mu\nu}
+\kappa^2 h^{\mu\lambda}h^\nu_\lambda
-\kappa^3 h^{\mu\lambda}h_{\lambda\alpha}h^{\alpha\nu}+\cdots~,
\end{equation}
and
\begin{equation}
\sqrt{-g}=\sqrt{-\det{g_{\mu\nu}}}
=1+\frac{\kappa}{2}h
+\frac{\kappa^2}{8}(h^2-2h^{\mu\nu}h_{\mu\nu})
+\frac{\kappa^3}{48}(h^3-6hh^{\mu\nu}h_{\mu\nu}+8h^\mu_\nu h^\nu_\lambda h^\lambda_\mu)+\cdots~,
\end{equation}
where $h^{\mu\nu}\equiv\eta^{\mu\alpha}h_{\alpha\beta}\eta^{\beta\nu}$ and
$h\equiv\eta^{\mu\nu}h_{\mu\nu}$.
%------------------------------------------------------------------------------
Using these expansions, we obtain 
%These expansions leads to
the Einstein-Hilbert action: 
%becomes
\begin{eqnarray}
\mathcal{L}_{EH}&=&\frac{1}{16\pi G}\sqrt{-g}R=\frac{2}{\kappa^2}\sqrt{-g}R\\
&=&\frac{1}{2}\left(\partial^\mu h^{\nu\lambda}\partial_\mu h_{\nu\lambda}
-\frac{1}{2}\partial^\mu h\partial_\mu h\right)%\\
%&&
+\kappa\left(\frac{1}{2}h^\alpha_\beta\partial^\mu h^\beta_\alpha\partial_\mu h
-\frac{1}{2}h^\alpha_\beta\partial_\alpha h^\mu_\nu\partial^\beta h^\nu_\mu
-h^\alpha_\beta\partial_\mu h^\nu_\alpha\partial^\mu h^\beta_\nu\right.\nonumber\\
%&&\qquad
&&\!\!+\left.
\frac{1}{4}h\partial^\alpha h^\mu_\nu\partial_\alpha h^\nu_\mu
+h^\beta_\mu\partial_\nu h^\alpha_\beta\partial^\mu h^\nu_\alpha
-\frac{1}{8}h\partial^\mu h\partial_\mu h\right)+\ldots~,
\end{eqnarray}
where we use the de Donder gauge, 
$\partial_\mu h^\mu_\nu=\frac{1}{2}\partial_\nu h$.
This expression involves a kinetic term as well as terms for an infinite number of interactions among gravitons.
%------------------------------------------------------------------------------
The Lagrangian of Maxwell theory coupled with gravitons becomes %reads
\begin{eqnarray}
\mathcal{L}_M&=&-\frac{\sqrt{-g}}{4}F^2
=-\frac{\sqrt{-g}}{4}g^{\alpha\beta}g^{\mu\nu}F_{\alpha\mu}F_{\beta\nu}\\
&=&-\frac{1}{4}\eta^{\alpha\beta}\eta^{\mu\nu}F_{\alpha\mu}F_{\beta\nu}
-\frac{\kappa}{2}h^{\mu\nu}
\left[-\eta^{\alpha\beta}F_{\alpha\mu}F_{\beta\nu}-
\eta_{\mu\nu}\left(-\frac{1}{4}\eta^{\gamma\delta}\eta^{\lambda\sigma}
F_{\gamma\lambda}F_{\delta\sigma}\right)\right]\nonumber\\
&&+\frac{\kappa^2}{4}
\left[\frac{1}{2}(h^2-2h^{\mu\nu}h_{\mu\nu})
\left(-\frac{1}{4}\eta^{\gamma\delta}\eta^{\lambda\sigma}
F_{\gamma\lambda}F_{\delta\sigma}\right)%\right.\\
%&&+\left.
F_{\alpha\beta}F_{\mu\nu}(hh^{\alpha\mu}\eta^{\beta\nu}
-2h^{\alpha\lambda}h^\mu_\lambda\eta^{\beta\nu}-h^{\alpha\mu}h^{\beta\nu})
\right]+\ldots ~,
\end{eqnarray}
where we choose the Lorenz gauge, $\partial^\mu A_\mu=0$.
%-----------------------------------------------------------------------------------------
We also find the coupling between the massive scalar boson and the dilaton as follows:
\begin{eqnarray}
\mathcal{L}_{KG}&=&
\mathcal{L}_{0}-a\phi(D_\mu\varphi)^*D^\mu\varphi
-a m^2\phi\varphi^*\varphi%+\ldots
-\frac{\kappa}{2}h^{\mu\nu}
\left(\partial_\mu\varphi^*\partial_\nu\varphi-\eta_{\mu\nu}\mathcal{L}_0\right)\nonumber\\
&&+\frac{\kappa^2}{2}\left[\frac{1}{4}(h^2-2h^{\mu\nu}h_{\mu\nu})\mathcal{L}_0
+\left(h^\mu_\lambda h^{\lambda\nu}-\frac{1}{2}hh^{\mu\nu}\right)
\partial_\mu\varphi^* \partial_\nu\varphi\right]
+\ldots ~,
\end{eqnarray}
where ${\cal L}_0$ is the Lagrangian in flat spacetime,
\begin{equation}
\mathcal{L}_{0}\equiv
\frac{1}{2}(\eta^{\mu\nu}\partial_\mu\varphi^*\partial_\nu\varphi-m^2\varphi^*\varphi)\,.
\end{equation}
%%%%%%%%%%%%%%%%%%%%%%%%%%%%%%%%%%%%%%%%%%%%%%%%%%%%%%%%%%%%%%%%%%%%%%%%%%%
%Elementary processes : gravitation
%%%%%%%%%%%%%%%%%%%%%%%%%%%%%%%%%%%%%%%%%%%%%%%%%%%%%%%%%%%%%%%%%%%%%%%%%%%
\section{Elementary processes : gravitation}
\begin{figure}[t]
\begin{center}
\includegraphics[width=12cm]{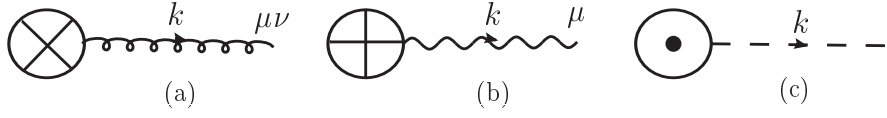}
\caption{The external field of graviton, photon and dilaton with momentum ${\bf k}$ 
%is shown 
in (a), (b) and (c), respectively. }
\label{propagators}
\end{center}
\end{figure}

Following Ref.~\cite{Paszko}, 
we use a classical source of gravitation:
$T^{\mu\nu}(x)=M\delta^\mu_0\delta^\nu_0\delta^{(3)}({\bf x})$.
According to the equation of motion, 
this source creates an external field $h^{ext}_{\mu\nu}({\bf k})$:
\begin{equation}
h_{\mu\nu}(k)=\frac{\kappa M}
{4{\bf k}^2}(\eta_{\mu\nu}-2\eta_{\mu 0}\eta_{\nu 0})2\pi\delta(k_0)
\equiv h^{ext}_{\mu\nu}({\bf k})2\pi\delta(k_0)~,
\end{equation}
where $k=p'-p$,~$k_0=0$ and $h^{ext}_{\mu\nu}({\bf k})$ is shown in Fig. \ref{propagators} (a).
%**************************************************************
%graviton external field no zu
%**************************************************************
We call amplitudes including the external field as ``semiclassical'' amplitudes.
%-----------------------------------------------------------------------------------------
The ``Newton'' potential is evaluated from the semiclassical amplitude 
including the graviton-scalar boson interaction.
%becomes,
%is as follows:
%for the first order, 
%Up to ${\cal O}(1/r)$,
%Up to the order of the inverse of the distance,
%the potential becomes
%**************************************************************
%iM^1 no zu (for graviton)
%**************************************************************
%\begin{equation}
%V(r)_{Newton1}=-\frac{GM}{r}\left(\frac{m^2+2{\bf p}^2}{E}\right)~,
%\end{equation}
%
%
%
%\begin{equation}
%i\mathcal{M}^{(1)}
%=ih^{ext}_{\alpha\beta}({\bf k})V^{\alpha\beta}(p',p)
%%&=&\frac{4\pi GM}{{\bf k}^2}(\eta_{\alpha\beta}-2\delta_\alpha^0\delta_\beta^0)
%%[p'^\alpha p^\beta+p'^\beta p^\alpha-\eta^{\alpha\beta}(p'\cdot p-m^2)]\\
%=-\frac{4\pi GM}{{\bf k}^2}(2m^2+4{\bf p}^2).
%\end{equation}
%\begin{equation}
%V(r)=\left.\int\frac{d^3\vk}{(2\pi)^3}e^{i{\bf k}\cdot{\bf r}}
%\frac{i\mathcal{M}^{(1)}}{\sqrt{2E'}\sqrt{2E}}\right|_{E'=E}%\\
%%&=&-4\pi GM\left(\frac{m^2+2{\bf p}^2}{E}\right)
%%\int\frac{d^3\vq}{(2\pi)^3}\frac{e^{i{\bf k}\cdot{\bf r}}}{{\bf k}^2}
%%=-\frac{GM}{r}\left(\frac{m^2+2{\bf p}^2}{E}\right)
%\end{equation}
%Newton potential ~$ V(r)=-\frac{GMm}{r}$~ for ${\bf p}=\v0$.
%
%
%
%and the second order,
Up to ${\cal O}(1/r^2)$,
the potential becomes
%**************************************************************
%iM^2 no zu (for graviton)
%**************************************************************
\begin{equation}
V(r)_{Newton}\approx
-\frac{GMm}{r}-\frac{3GM{\bf p}^2}{2mr}+\frac{G^2M^2m}{2r^2}+\cdots 
\,.\label{Newton2}
\end{equation}
%%
%%
%%\begin{eqnarray}
%%V(r) &\approx& -\frac{GMm}{r}-\frac{3GM{\bf p}^2}{2mr}+\frac{G^2M^2m}{2r^2}+\cdots \\
%%&=&-\frac{GMm}{r}+\frac{G^2M^2m}{2r^2}+\cdots 
%%~~~(for ~~{\bf p}=\v0)
%%\end{eqnarray}
%%
%%
%\begin{eqnarray}
%V(r)
%&=&
%\left.\int\frac{d^3\vk}{(2\pi)^3}e^{i{\bf k}\cdot{\bf r}}
%\left[\frac{i\mathcal{M}^{total}}{\sqrt{4E'E}}
%-\int\frac{d^3p''}{(2\pi)^3}
%\frac{\frac{i\mathcal{M}^{(1)}_G({\bf p}',{\bf p}'')}
%{\sqrt{4E'E''}}\frac{i\mathcal{M}^{(1)}_G({\bf p}'',{\bf p})}{\sqrt{4E''E}}}{E-E''}
%\right]\right|_{E'=E}\\
%&\approx&
%-\frac{GMm}{r}-\frac{3GM{\bf p}^2}{2mr}+\frac{G^2M^2m}{2r^2}+\cdots
%\end{eqnarray}
%For ${\bf p}=\v0$,
%\begin{equation}
%V(r)=-\frac{GMm}{r}+\frac{G^2M^2m}{2r^2}+\cdots
%\end{equation}
The first term in (\ref{Newton2}) is recognized as the attractive Newton potential.
%%%%%%%%%%%%%%%%%%%%%%%%%%%%%%%%%%%%%%%%%%%%%%%%%%%%%%%%%%%%%%%%%%%%%%%%%%%
%Elementary process : electromagnetism
%%%%%%%%%%%%%%%%%%%%%%%%%%%%%%%%%%%%%%%%%%%%%%%%%%%%%%%%%%%%%%%%%%%%%%%%%%%
\section{Elementary process : electromagnetism}
We use a static charge as a classical source:
$J^{\mu}(x)=Q\delta^\mu_0\delta^{(3)}({\bf x})$,
as usual.
This source creates an external field $A^{ext} _\mu$: 
\begin{equation}
A^{\mu}(k)=\frac{Q}{{\bf k}^2}\delta^\mu_02\pi\delta(k_0)
\equiv A^{ext ~\mu}({\bf k})2\pi\delta(k_0)~,
\end{equation}
where $k=p'-p$,~$k_0=0$ and $A^{ext ~\mu}$ is %an external field as 
shown in Fig. \ref{propagators} (b).
%**************************************************************
%photon external field no zu
%**************************************************************
% 
%**************************************************************
%iM^1 no zu (for photon)
%**************************************************************
We obtain ``Coulomb'' potential
from the semiclassical amplitude,
where several interactions 
between photons, scalar bosons and gravitons, 
as well as between photons and scalar bosons, 
are included.
Up to ${\cal O}(1/r^2)$,
the potential becomes
%Coulomb potential $V(r)_{Coulomb1}=\frac{Qq}{4\pi r}$ is obtained
%from the semiclassical amplitude including the photon - scalar boson interaction
%for the first order.
%Note that there is no dependence on $\vp$ in $V(r)_{Coulomb1}$.
%
%We obtain Coulomb potential $V(r)_{Coulomb1}=\frac{Qq}{4\pi r}$
%by evaluating from the semiclassical amplitude including the photon - scalar boson interaction
%for the first order.
%
%\begin{equation}
%i\mathcal{M}^{(1)}_e=
%iA^{ext}_{\alpha}({\bf k})V^{\alpha}(p',p)
%%=\frac{Qq}{{\bf k}^2}\, \eta_{\alpha 0}[p^\alpha+p'^\alpha]
%=\frac{Qq}{{\bf k}^2}(2E).
%\end{equation}
%\begin{equation}
%V(r)=\left.\int\frac{d^3\vk}{(2\pi)^3}e^{i{\bf k}\cdot{\bf r}}
%\frac{i\mathcal{M}^{(1)}_e}{\sqrt{2E'}\sqrt{2E}}\right|_{E'=E}
%=Qq\int\frac{d^3\vq}{(2\pi)^3}\frac{e^{i{\bf k}\cdot{\bf r}}}{{\bf k}^2}
%=\frac{Qq}{4\pi r}
%\end{equation}
%Coulomb potential ~$V(r)=\frac{Qq}{4\pi r}$.
%Note that there is no dependence on $\vp$.
%**************************************************************
%iM^2 no zu (for photon)
%**************************************************************
%For the second order of the semiclassical amplitude,
%which includes the several interactions 
%between photon, scalar boson and graviton, 
%as well as between photon and scalar boson, 
%the ``Coulomb'' potential is evaluated as follows:
%evaluated from the semiclassical amplitude including these interaction 
%becomes 
\begin{equation}
V(r)_{Coulomb} \approx
\frac{Qq}{4\pi r}+\frac{GmQ^2}{8\pi r^2}-\frac{GMQq}{4\pi r^2}+\cdots 
\,.\label{Coulomb2}
\end{equation}
%where several interactions 
%between photon, scalar boson and graviton, 
%as well as between photon and scalar boson, 
%are included in the amplitude.
%
Note that the first term in (\ref{Coulomb2}) is the Coulomb potential
%and there is 
with no dependence on the momentum $\vp$. %in $V(r)_{Coulomb1}$.

%
%\begin{eqnarray}
%V(r)_{Coulomb2}
%&=&\left.\int\frac{d^3\vk}{(2\pi)^3}e^{i{\bf k}\cdot{\bf r}}
%\left[\frac{i\mathcal{M}^{subtotal}}{2E}
%-2\int\frac{d^3p''}{(2\pi)^3}
%\frac{\frac{i\mathcal{M}^{(1)}_e({\bf p},{\bf p}'')}{\sqrt{4EE''}}
%\frac{i\mathcal{M}^{(1)}_G({\bf p}'',{\bf p})}{\sqrt{4E''E}}}{E-E''}
%\right]\right.\\
%&\approx&
%\frac{Qq}{4\pi r}+\frac{GmQ^2}{8\pi r^2}-\frac{GMQq}{4\pi r^2}+\cdots~.
%\end{eqnarray}
%%%%%%%%%%%%%%%%%%%%%%%%%%%%%%%%%%%%%%%%%%%%%%%%%%%%%%%%%%%%%%%%%%%%%%%%%%%
%Elementary process : Dilaton
%%%%%%%%%%%%%%%%%%%%%%%%%%%%%%%%%%%%%%%%%%%%%%%%%%%%%%%%%%%%%%%%%%%%%%%%%%%
\section{Elementary process : Dilaton}
\begin{figure}[t]
\begin{center}
\includegraphics[width=14cm]{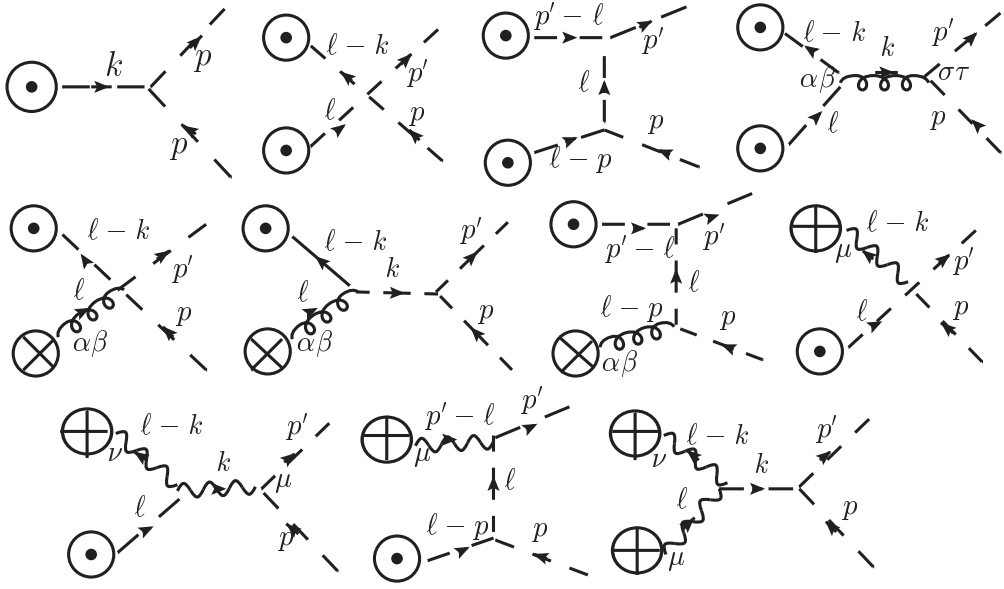}
\caption{The diagrams for the first and second order of interactions 
between gravitons, photons and dilatons.
These diagrams include the classical sources shown in Fig. \ref{propagators}.}
%the dilatonic force
%the external fields
%the classical sources
\label{dilaton1st2nd}
\end{center}
\end{figure}

Now, we introduce a dilaton field
%and use
%Using
and employ 
a dilatonic charge
$\rho_{\Sigma}=\Sigma\,\delta^{(3)}(\vx)$,
%is employed
as a classical source.
This source 
creates an external field $\phi^{ext}(\vk)$,
which is shown in Fig. \ref{propagators} (c):
%**************************************************************
%Dilaton external field no zu
%**************************************************************
\begin{equation}
\phi^{ext}(\vk)\equiv -\frac{aM}{\vk^2}\equiv -\frac{\Sigma}{\vk^2}~.
\end{equation}
%**************************************************************
%iM^1 no zu (for Dilaton)
%**************************************************************
%We evaluate the semiclassical amplitudes including 
%interactions between gravitons, scalar bosons and dilatons.
%
%
%Since the amplitude of one intermediating dilaton,
%which is shown in Fig.,
%is found to be
%\begin{equation}
%i\mathcal{M}^{(1)}_d
%=-a^2M\, \frac{1}{\vk^2}(EE'-\vp\cdot(\vp+\vk)+m^2)
%\approx -a^2M\,\frac{2m^2}{\vk^2}\,,
%\end{equation}
%the potential of the dilatonic force can be read as
%\begin{equation}
%V(r)_{dilaton1}=-\frac{a^2Mm}{4\pi r}\left(\frac{m}{E}\right)
%\approx -\frac{a^2Mm}{4\pi r}\left(1-\frac{\vp^2}{2m^2}\right)\,.
%\end{equation}
%
%
%For ${\bf p}={\bf 0}$,
%\begin{equation}
%V(r) \approx -\frac{a^2Mm}{4\pi r}\,.
%\end{equation}
%**************************************************************
%iM^2 no zu (for Dilaton)
%**************************************************************
%
%The succeeding diagrams for the second order of interactions
The diagrams for the first and second order of interactions
are gathered in Fig. \ref{dilaton1st2nd}.
These amplitudes include 
the gravitational, the electrical and the dilatonic sources.

We finally evaluate the semiclassical amplitudes of the scalar boson including 
the dilatonic force and interactions
between gravitons, photons and dilatons.
% 
%including interactions with dilaton fields are figured in Fig..
%These amplitudes with the gravitational, the electrical and the dilatonic sources are figured in Fig..
%
%For the second order of interactions,
%We evaluate amplitudes with the gravitational, the electrical and the dilatonic sources
%for the second order.
%
%
%
%
The sum of the contributions of the diagrams gives the additional potential due to the dilatonic interaction.
The potential of the dilatonic force can be read as
\begin{equation}
V (r)_{dilaton}\approx
\!\! {-\frac{a^2Mm}{4\pi r}+\frac{a^2M\vp^2}{8\pi m r}+\frac{a^2Q^2m}{2(4\pi)^2 r^2}+\frac{a^2M^2m}{(4\pi)^2r^2}
-\frac{a^2MQq}{(4\pi)^2r^2} +\frac{a^4M^2m}{2(4\pi)^2r^2}}\,\label{dilaton2}.
\end{equation}

%\begin{eqnarray}
%V (r)_{dilaton2}&=&\left.\int\frac{d^3\vk}{(2\pi)^3}e^{i{\bf k}\cdot{\bf r}}
%\left[\frac{i\mathcal{M}^{subtotal}}{2E}
%-\sum\int\frac{d^3p''}{(2\pi)^3}
%\frac{
%\frac{i\mathcal{M}^{(1)}_A({\bf p},{\bf p}'')}{\sqrt{4EE''}}
%\frac{i\mathcal{M}^{(1)}_B({\bf p}'',{\bf p})}{\sqrt{4E''E}}}{E-E''}\right]\right.\\
%&\approx&
%\!\! {-\frac{a^2Mm}{4\pi r}+\frac{a^2M\vp^2}{8\pi m r}+\frac{a^2Q^2m}{2(4\pi)^2 r^2}+\frac{a^2M^2m}{(4\pi)^2r^2}
%-\frac{a^2MQq}{(4\pi)^2r^2} +\frac{a^4M^2m}{2(4\pi)^2r^2}}\,\label{dilaton2}.
%\end{eqnarray}
%where
%$(A, B)=(G,d), (d,G), (e,d), (d,e), (d,d)$,
%
%For  ${\bf p}=\v0$,
%\begin{equation}
%V (r)\approx
%-\frac{a^2Mm}{4\pi r}+\frac{a^2Q^2m}{2(4\pi)^2 r^2}+\frac{a^2M^2m}{(4\pi)^2r^2}
%-\frac{a^2MQq}{(4\pi)^2r^2} +\frac{a^4M^2m}{2(4\pi)^2r^2}\,.
%\end{equation}
%%%%%%%%%%%%%%%%%%%%%%%%%%%%%%%%%%%%%%%%%%%%%%%%%%%%%%%%%%%%%%%%%%%%%%%%%%%
%Cancellation of potential
%%%%%%%%%%%%%%%%%%%%%%%%%%%%%%%%%%%%%%%%%%%%%%%%%%%%%%%%%%%%%%%%%%%%%%%%%%%
\section{Cancellation of potential}
We consider a static case.
If we set $\vp=\v0$ and impose balance conditions,
 $Q=\sqrt{1+a^2}M,\,q=\sqrt{1+a^2}m$,
in (\ref{Newton2}), (\ref{Coulomb2}) and (\ref{dilaton2}), %zero momentum
the static potentials are cancelled each other
up to ${\cal O}(1/r^2)$:
%Cancellation of the static potential.
\begin{equation}
V_{Newton} + V_{Coulomb} + V_{dilaton} =0\,.
\end{equation}
%
%For ${\bf p}={\bf 0}$,
%\begin{equation}
%V(r) \approx -\frac{a^2Mm}{4\pi r}\,.
%\end{equation}
%For  ${\bf p}=\v0$,
%\begin{equation}
%V (r)\approx
%-\frac{a^2Mm}{4\pi r}+\frac{a^2Q^2m}{2(4\pi)^2 r^2}+\frac{a^2M^2m}{(4\pi)^2r^2}
%-\frac{a^2MQq}{(4\pi)^2r^2} +\frac{a^4M^2m}{2(4\pi)^2r^2}\,.
%\end{equation}
%
%
%%%%%%%%%%%%%%%%%%%%%%%%%%%%%%%%%%%%%%%%%%%%%%%%%%%%%%%%%%%%%%%%%%%%%%%%%%%
%Summary and Outlook
%%%%%%%%%%%%%%%%%%%%%%%%%%%%%%%%%%%%%%%%%%%%%%%%%%%%%%%%%%%%%%%%%%%%%%%%%%%
\section{Summary and Outlook}
We evaluated the potential of long-range forces coupled with the dilaton field from Feynman diagrams.
Up to ${\cal O}(1/r^2)$, 
we showed cancellation of the static potential under the balance condition, $Q=\sqrt{1+a^2}M$.
In future, we wish to study the following subjects;  
%Our future works are as follows:\\
higher-derivative theories, %(includes dimensionful constants) \\
calculation on the classical curved background which realizes a three-body system,
loop corrections as in Ref.~\cite{Faller,BB}, %(Quantum Moduli Space of two or more "particles")\\
higher-dimensional theories, %($1/D$-expansion ?)\\
$\cdots$ and much more !
%%%%%%%%%%%%%%%%%%%%%%%%%%%%%%%%%%%%%%%%%%%%%%%%%%%%%%%%%%%%%%%%%%%%%%%%%%%
%Acknowledgements
%%%%%%%%%%%%%%%%%%%%%%%%%%%%%%%%%%%%%%%%%%%%%%%%%%%%%%%%%%%%%%%%%%%%%%%%%%%
\section*{Acknowledgements}
The authors would like to thank K. Shinoda for comments, and also the organizers of JGRG18.
%%%%%%%%%%%%%%%%%%%%%%%%%%%%%%%%%%%%%%%%%%%%%%%%%%%%%%%%%%%%%%%%%%%%%%%%%%%
%Figures
%%%%%%%%%%%%%%%%%%%%%%%%%%%%%%%%%%%%%%%%%%%%%%%%%%%%%%%%%%%%%%%%%%%%%%%%%%%
%%%%%%%%%%%%%%%%%%%%%%%%%%%%%%%%%%%%%%%%%%%%%%%%%%%%%%%%%%%%%%%%%%%%%%%%%%%
%thebibliography
%%%%%%%%%%%%%%%%%%%%%%%%%%%%%%%%%%%%%%%%%%%%%%%%%%%%%%%%%%%%%%%%%%%%%%%%%%%


\begin{thebibliography}{99}
\bibitem{MP} S.~D.~Majumdar, Phys. Rev. {\bf 72} (1947) 930;
A.~Papapetrou, Proc. R. Irish Acad. {\bf A51} (1947) 191.

\bibitem{KSJMP} K.~Shiraishi, J. Math. Phys. {\bf 34} (1993) 1480.

\bibitem{DS} Y.~Degura and K.~Shiraishi, Class. Q. Grav. {\bf 17} (2000) 4031.

\bibitem{Paszko} 
R.~Paszko, arXiv:0801.1835v2 [gr-qc]

\bibitem{Faller} S.~Faller, Phys. Rev. {\bf D77} (2008) 124039.

\bibitem{BB} N.~E.~J.~Bjerrum-Bohr, hep-th/0206236v3 (2007).

%\bibitem{AB1} G.~Alekseev and V.~Belinski, Phys. Rev. {\bf D76} (2007) R021501.
%\bibitem{AB2} G.~Alekseev and V.~Belinski, ArXiv: 0710.2515v1 [gr-qc].
%\bibitem{PP} M.~Pizzi and A.~Paolino, Int. J. Mod. Phys. {\bf A23} (2008) 1222.

%\bibitem{Iwasaki} Y.~Iwasaki, Prog. Theor. Phys. {\bf 46} (1971) 1587.



%\bibitem{BO1} B.~M.~Baker and R.~F.~O'Connell, Phys. Rev. {\bf D12} (1975) 329.
%\bibitem{BO2} B.~M.~Baker and R.~F.~O'Connell, J. Math. Phys. {\bf 18} (1977) 1818; 
%{\t ibid.} {\bf 19} (1978) 1231.
%\bibitem{BO3} B.~M.~Baker and R.~F.~O'Connell, Lett Nuovo Cim. {\bf 19} (1977) 467.


%\bibitem{SM} K.~Shiraishi and T.~Maki, Phys. Rev. {\bf D53} (1996) 3070.
%\bibitem{KMS} N.~Kan, T.~Maki and K.~Shiraishi, Phys. Rev. {\bf D64} (2001) 104009.

 
\end{thebibliography}
\end{document}